\documentclass[twocolumn,nofootinbib]{revtex4-1}
\usepackage{amsmath,amssymb,bm,graphicx,hyperref}
\allowdisplaybreaks[1]

\renewcommand\d{\partial}
\newcommand\+{\dagger}

\renewcommand\>{\rangle}
\newcommand\p{\bm{p}}
\newcommand\x{\bm{x}}

\begin{document}

\title{Universal bound states of one-dimensional bosons\\with two- and three-body attractions}

\author{Yusuke Nishida}
\affiliation{Department of Physics, Tokyo Institute of Technology,
Ookayama, Meguro, Tokyo 152-8551, Japan}

\date{March 2017}

\begin{abstract}
When quantum particles are confined into lower dimensions, an effective three-body interaction inevitably arises and may cause significant consequences.
Here we study bosons in one dimension with weak two-body and three-body interactions, predict the existence of two three-body bound states when both interactions are attractive, and determine their binding energies as universal functions of the two-body and three-body scattering lengths.
We also show that an infinitesimal three-body attraction induces an excited bound state only for 3, 39, or more bosons.
Our findings herein have direct relevance to a broad range of quasi-one-dimensional systems realized with ultracold atoms.
\end{abstract}

\maketitle

\section{Introduction}
Effective three- and higher-body interactions are ubiquitous and play important roles in various subfields of physics~\cite{Primakoff:1939,Axilrod:1943,Muto:1943,Fujita:1957,Hammer:2013}.
One such example is provided by quantum particles confined into lower dimensions even when their interaction in free space is purely pairwise.
As far as low-energy physics relative to the transverse excitation energy is concerned, the system admits an effective low-dimensional description where multibody interactions inevitably arise from virtual transverse excitations.
In particular, the three-body interaction in one-dimensional systems may cause significant consequences because it breaks the integrability~\cite{Muryshev:2002,Sinha:2006,Mazets:2008} and is marginally relevant when attractive~\cite{Sekino:2018,Drut:2018}.
The purpose of this work is to elucidate possible consequences of the three-body interaction for bound states of bosons in one dimension.

\subsection{Model and universality}
Bosons in one dimension with two-body and three-body interactions are described by
\begin{align}
H = \int\!dx\left[\frac1{2m}\frac{d\phi^\+(x)}{dx}\frac{d\phi(x)}{dx}
+ \frac{u_2}{2m}|\phi(x)|^4 + \frac{u_3}{6m}|\phi(x)|^6\right],
\end{align}
where we set $\hbar=1$ and $|\phi(x)|^{2n}\equiv[\phi^\+(x)]^n[\phi(x)]^n$.
When this system is realized by confining weakly interacting bosons with a two-dimensional harmonic potential~\cite{Bloch:2008}, the two-body and three-body couplings are provided by
\begin{align}\label{eq:coupling}
u_2 = 2\frac{a_\mathrm{3D}}{l_\perp^2}
\qquad\text{and}\qquad
u_3 = -12\ln(4/3)\frac{a_\mathrm{3D}^2}{l_\perp^2},
\end{align}
respectively, for $|a_\mathrm{3D}|\ll l_\perp$, where $a_\mathrm{3D}$ is the $s$-wave scattering length in free space and $l_\perp\equiv1/\sqrt{m\omega_\perp}$ is the harmonic oscillator length~\cite{Olshanii:1998,Tan:2010}.%
\footnote{Our result for $u_3$ is four times smaller than that in Refs.~\cite{Sinha:2006,Mazets:2008} but agrees with Ref.~\cite{Tan:2010}.}
While the two-body interaction can be either attractive or repulsive depending on the sign of $a_\mathrm{3D}$, the three-body interaction is always attractive ($u_3<0$) because it arises from the second-order perturbation theory~\cite{Mazets:2008}.
We note that four- and higher-body interactions also exist but are irrelevant to low-energy physics.

It is more convenient to parametrize the two-body and three-body couplings in terms of the scattering lengths.
The two-body scattering length is introduced as $a_2\equiv-2/u_2$.
With this definition, the binding energy of a two-body bound state (dimer) is provided by $E_2=-1/(ma_2^2)$ for $a_2\gg l_\perp$~\cite{Bloch:2008}.
Similarly, the three-body scattering length is introduced so that the binding energy of a three-body bound state (trimer) is provided by $E_3\equiv-1/(ma_3^2)$ for $a_3\gg l_\perp$ when the two-body interaction is assumed to be absent~\cite{Sekino:2018}.
This definition leads to $a_3\sim e^{-\sqrt3\pi/u_3}l_\perp$ as we will see later in Eq.~(\ref{eq:scattering_length}).
While $a_3\gg|a_2|\gg l_\perp$ is naturally realized for weakly interacting bosons with $|a_\mathrm{3D}|\ll l_\perp$, we study the system with an arbitrary $-\infty<a_3/a_2<+\infty$ because the two-body and three-body interactions are independently tunable in principle with ultracold atoms~\cite{Daley:2014,Petrov:2014a,Petrov:2014b,Paul:2016}.
As far as both interactions are weak in the sense of $|a_2|,a_3\gg l_\perp$, low-energy physics of the system at $|E|\ll1/(ml_\perp^2)$ is universal, i.e., depends only on the two scattering lengths.

\section{Three-boson system}
\subsection{Formulation}
We now focus on the system of three bosons whose Schr\"odinger equation reads
\begin{align}
\left[-\frac1{2m}\sum_{i=1}^3\frac{\d^2}{\d x_i^2} + \frac{u_2}{m}\sum_{1\leq i<j\leq3}\delta(x_{ij})
+ \frac{u_3}{m}\delta(x_{12})\delta(x_{23})\right] \notag\\
{} \times \Psi(x_1,x_2,x_3) = E\,\Psi(x_1,x_2,x_3),
\end{align}
where $x_{ij}\equiv x_i-x_j$ is the interparticle separation.
For a bound state with its binding energy $E\equiv-\kappa^2/m<0$, the Schr\"odinger equation is formally solved in Fourier space by
\begin{align}\label{eq:solution}
\tilde\Psi(p_1,p_2,p_3) = -\frac{\sum_{i=1}^3\tilde\Psi_2(P_{123}-p_i;p_i)
+ \tilde\Psi_3(P_{123})}{\kappa^2+\sum_{i=1}^3\frac{p_i^2}{2}},
\end{align}
where $P_{123}\equiv p_1+p_2+p_3$ is the center-of-mass momentum and
\begin{subequations}\label{eq:Psi}
\begin{align}
\tilde\Psi_2(P;p) &\equiv u_2\int\!\frac{dq}{2\pi}\,\tilde\Psi(P-q,q,p), \\
\tilde\Psi_3(P) &\equiv u_3\int\!\frac{dq\,dr}{(2\pi)^2}\,\tilde\Psi(P-q-r,q,r)
\end{align}
\end{subequations}
are the Fourier transforms of $u_2\Psi(X,X,x)$ and $u_3\Psi(X,X,X)$, respectively.
After rewriting $p_1\to P-p-q$, $p_2\to p$, and $p_3\to q$ in Eq.~(\ref{eq:solution}), the integration over $q$ leads to
\begin{subequations}\label{eq:coupled_Psi}
\begin{align}
\frac1{u_2}\tilde\Psi_2(P-p;p)
= -\int\!\frac{dq}{2\pi}\frac{2\tilde\Psi_2(P-q;q)}{\kappa^2+\frac{(P-p-q)^2+p^2+q^2}{2}} \notag\\
{} - \frac{\tilde\Psi_2(P-p;p) + \tilde\Psi_3(P)}{2\sqrt{\kappa^2+\frac{(P-p)^2}{4}+\frac{p^2}{2}}},
\end{align}
while the integration over $p$ and $q$ leads to
\begin{align}
\frac1{u_3}\tilde\Psi_3(P)
= -\int\!\frac{dq}{2\pi}\frac{3\tilde\Psi_2(P-q;q)}{2\sqrt{\kappa^2+\frac{(P-q)^2}{4}+\frac{q^2}{2}}} \notag\\
{} - \frac1{\sqrt3\pi}\ln\!\left(\frac{\Lambda}{\sqrt{\kappa^2+\frac{P^2}{6}}}\right)\tilde\Psi_3(P),
\end{align}
\end{subequations}
where $\Lambda\sim l_\perp^{-1}$ is the momentum cutoff and Eqs.~(\ref{eq:Psi}) are used on the left-hand sides.
Finally, by substituting the ansatz of $\tilde\Psi_2(P-p;p)\equiv2\pi\delta(P)\tilde\psi_2(p)$ and $\tilde\Psi_3(P)\equiv2\pi\delta(P)\tilde\psi_3$ (i.e., zero center-of-mass momentum) into Eqs.~(\ref{eq:coupled_Psi}) as well as the two-body and three-body couplings parametrized as
\begin{align}\label{eq:scattering_length}
u_2 = -\frac2{a_2}
\qquad\text{and}\qquad
u_3 = -\frac{\sqrt3\pi}{\ln(a_3\Lambda)},
\end{align}
we obtain
\begin{subequations}\label{eq:coupled_psi}
\begin{align}\label{eq:two-body}
& \left(\frac{a_2}{2} - \frac1{2\sqrt{\kappa^2+\frac{3p^2}{4}}}\right)\tilde\psi_2(p) \notag\\
&= \int\!\frac{dq}{2\pi}\frac{2\tilde\psi_2(q)}{\kappa^2+p^2+q^2+pq}
+ \frac{\tilde\psi_3}{2\sqrt{\kappa^2+\frac{3p^2}{4}}}
\end{align}
and
\begin{align}\label{eq:three-body}
\frac{\ln(a_3\kappa)}{\sqrt3\pi}\tilde\psi_3
= \int\!\frac{dq}{2\pi}\frac{3\tilde\psi_2(q)}{2\sqrt{\kappa^2+\frac{3q^2}{4}}}.
\end{align}
\end{subequations}
Equation~(\ref{eq:two-body}) with $\tilde\psi_3$ eliminated by Eq.~(\ref{eq:three-body}) provides the closed one-dimensional integral equation for $\tilde\psi_2(p)$, which is to be solved numerically.
We note that nontrivial solutions exist only in the even-parity channel where $\tilde\psi_2(p)=\tilde\psi_2(-p)$.

As we can see in Eq.~(\ref{eq:scattering_length}), the positive (negative) two-body scattering length corresponds to the attractive (repulsive) two-body interaction.
The two-body attraction increases with increasing $1/a_2$ from the strong repulsion $1/a_2\to-\infty$ via no interaction $1/a_2=0$ to the strong attraction $1/a_2\to+\infty$.
On the other hand, the three-body scattering length is positive definite and the three-body attraction increases with increasing $1/a_3$ from the weak attraction $1/a_3\to+0$ to the strong attraction $1/a_3\to+\infty$.
For later discussion, we identify the prefactor of $\tilde\psi_3$ in Eq.~(\ref{eq:three-body}) as $-1/\bar u_3(\kappa)$, where
\begin{align}\label{eq:renormalized}
\bar u_3(\kappa) \equiv -\frac{\sqrt3\pi}{\ln(a_3\kappa)}
\end{align}
is the renormalized three-body coupling with logarithmic energy dependence~\cite{Sekino:2018}.

\subsection{Binding energies}
\begin{figure}[t]
\includegraphics[width=0.9\columnwidth,clip]{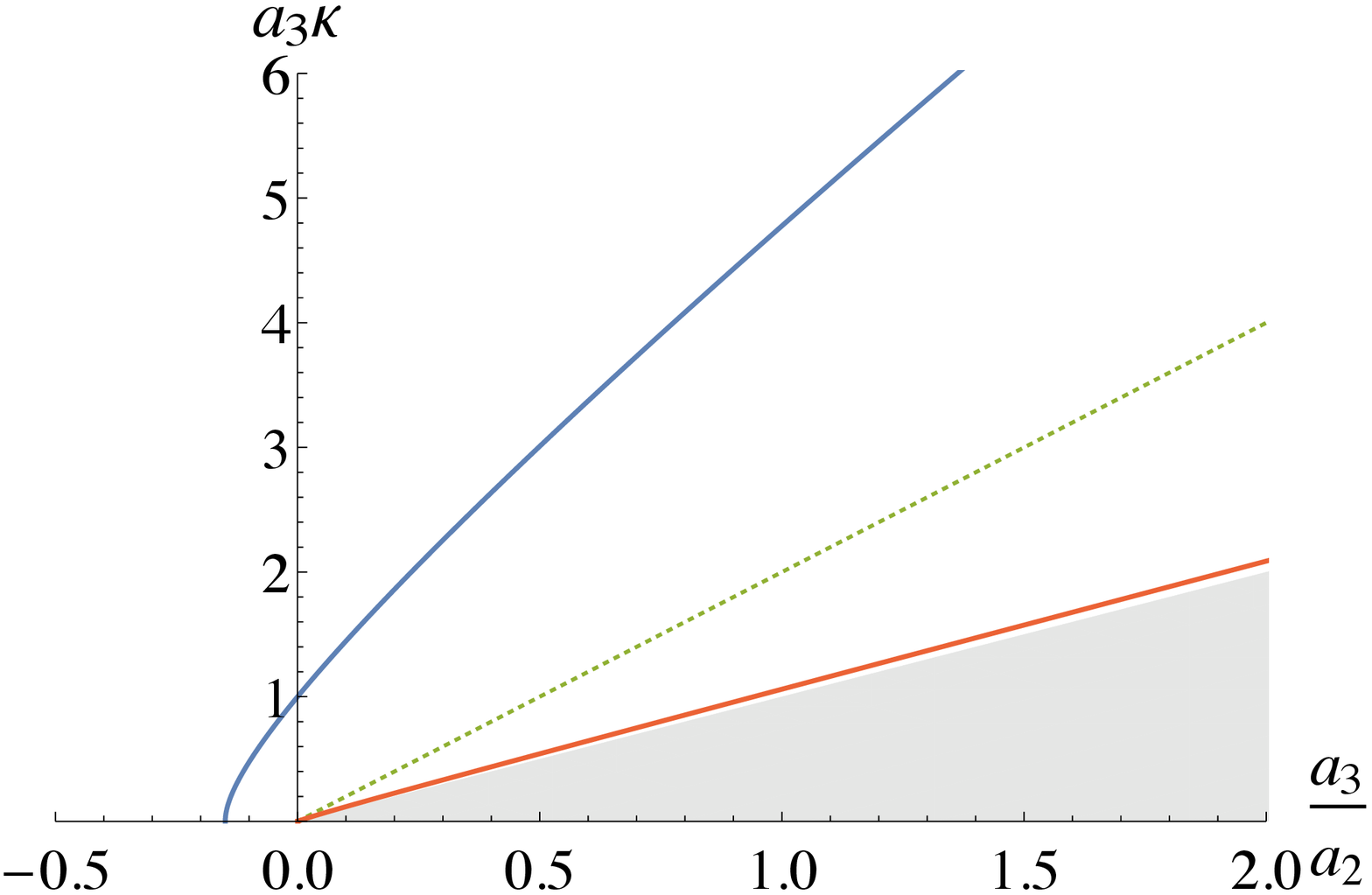}
\includegraphics[width=0.9\columnwidth,clip]{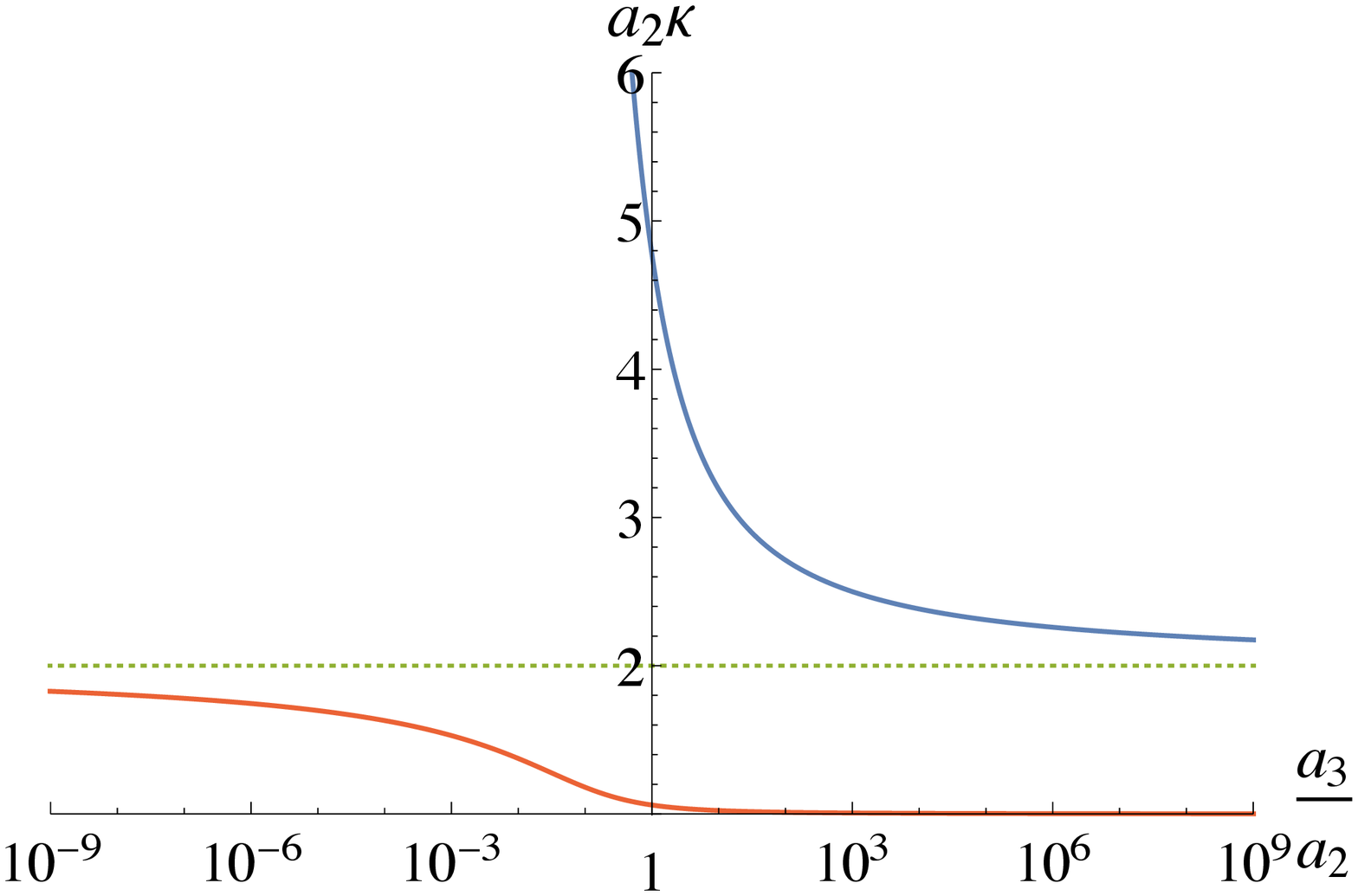}
\caption{\label{fig:a32kappa}
Binding energies of three-body bound states $E=-\kappa^2/m$ in the forms of $a_3\kappa$ (top panel) and $a_2\kappa$ (bottom panel) as functions of the three-body to two-body scattering length ratio $a_3/a_2$.
The upper (lower) solid curve corresponds to the ground (excited) state and the dotted line indicates $\kappa=2/a_2$ for the McGuire trimer.
The shaded region in the top panel indicates the atom-dimer continuum where $\kappa<\theta(a_2)/a_2$.}
\end{figure}

The numerical solutions for $\kappa>\theta(a_2)/a_2$ are plotted as functions of $a_3/a_2$ in Fig.~\ref{fig:a32kappa} with the different normalizations.%
\footnote{Their analytical expressions were recently obtained in Ref.~\cite{Guijarro:2018}.}
Here we find that the ground state trimer appears at $a_3/a_2\approx-0.149218$.
Its binding energy is $\kappa=1/a_3$ at $a_3/a_2=0$ by the definition of $a_3$ and asymptotically approaches $\kappa=2/a_2$ as
\begin{align}\label{eq:ground_weak}
\kappa \to \frac2{a_2} + \frac{2\pi}{\sqrt3\,a_2\ln(a_3/a_2)}
\qquad\text{toward}\qquad
\frac{a_3}{a_2} \to +\infty.
\end{align}
On the other hand, we find that the excited state trimer appears right at $a_3/a_2=0$ where the dimer state also appears.
Its binding energy asymptotically approaches $\kappa=2/a_2$ as
\begin{align}\label{eq:excited_strong}
\kappa \to \frac2{a_2} + \frac{2\pi}{\sqrt3\,a_2\ln(a_3/a_2)}
\qquad\text{toward}\qquad
\frac{a_3}{a_2} \to +0,
\end{align}
while it asymptotically approaches $\kappa=1/a_2$ as
\begin{align}\label{eq:excited_weak}
\kappa \to \frac1{a_2} + \frac{\pi^2}{18\,a_2\ln^2(a_3/a_2)}
\qquad\text{toward}\qquad
\frac{a_3}{a_2} \to +\infty.
\end{align}
The subleading term in Eq.~(\ref{eq:excited_weak}) indicates that the atom-dimer scattering length is provided by $\alpha_{1,2}\to3\sqrt3\,a_2\ln(a_3/a_2)/(2\pi)\gg a_2$ for $\ln(a_3/a_2)\to+\infty$.
This is consistent with the one obtained from the expectation value of the renormalized three-body interaction energy $V_3=[\bar u_3(\kappa)/m]\delta(x_{12})\delta(x_{23})$ with respect to the wave function right at the atom-dimer threshold; $\Psi(x_1,x_2,x_3)=\sqrt{\frac1{3a_2L^2}}\left[\sum_{1\leq i<j\leq3}e^{-|x_{ij}|/a_2}-4e^{-\sum_{1\leq i<j\leq3}|x_{ij}|/(2a_2)}\right]$~\cite{Kartavtsev:2009}.
We note that the wave functions here and below are all normalized on a line of length $L\gg a_2$.

When the three-body interaction is assumed to be absent, McGuire predicted a single trimer state with its binding energy $\kappa=2/a_2$~\cite{McGuire:1964}.
We find above that an infinitesimal three-body attraction immediately induces another trimer state appearing from the atom-dimer threshold at $\kappa=1/a_2$ as in Eq.~(\ref{eq:excited_weak}).
While our ground state trimer unsurprisingly reduces to the McGuire trimer in the limit of strong two-body or weak three-body attraction [Eq.~(\ref{eq:ground_weak})], it is interesting that our excited state trimer also reduces to the McGuire trimer in the opposite limit of weak two-body or strong three-body attraction [Eq.~(\ref{eq:excited_strong})].
This is because the renormalized three-body coupling in Eq.~(\ref{eq:renormalized}) turns out to be positive and vanishingly small toward the three-boson threshold $a_3\kappa\to+0$.
Indeed, the subleading terms in Eqs.~(\ref{eq:ground_weak}) and (\ref{eq:excited_strong}) for $\ln(a_3/a_2)\to\pm\infty$ can both be obtained from the expectation value of the renormalized three-body interaction energy $V_3=[\bar u_3(\kappa)/m]\delta(x_{12})\delta(x_{23})$ with respect to the wave function of the McGuire trimer; $\Psi(x_1,x_2,x_3)=\sqrt{\frac{8}{3a_2^2L}}e^{-\sum_{1\leq i<j\leq3}|x_{ij}|/a_2}$~\cite{Castin:2001}.

\section{\boldmath$N$-boson system}
While we have so far focused on the system of three bosons, it is straightforward to generalize our formulation and some results to an arbitrary $N$ number of bosons.
In particular, when the three-body interaction is assumed to be absent, McGuire also predicted a single $N$-body bound state for every $N$ with its binding energy $E_N^\mathrm{(MG)}\equiv-N(N^2-1)/(6ma_2^2)$~\cite{McGuire:1964}.
Its wave function in the domain of $x_1<x_2<\dots<x_N$ is provided by
\begin{align}\label{eq:bound-state}
\Psi_N(\x) = \sqrt{\frac{(N-1)!}{NL}\left(\frac2{a_2}\right)^{N-1}}
\exp\!\left(\sum_{i=1}^N\frac{N+1-2i}{a_2}x_i\right),
\end{align}
where $\x\equiv(x_1,x_2,\dots,x_N)$~\cite{Castin:2001}.
Then, the expectation value of the renormalized three-body interaction energy $V_3=[\bar u_3(\kappa)/m]\sum_{1\leq i<j<k\leq N}\delta(x_{ij})\delta(x_{jk})$ with respect to the wave function in Eq.~(\ref{eq:bound-state}) leads to the binding-energy shift induced by an infinitesimal three-body attraction, which is found to be
\begin{align}\label{eq:ground_N-body}
\Delta E_N \equiv E_N - E_N^\mathrm{(MG)}
\to -\frac{\sqrt3\pi N(N^2-1)(N^2-4)}{45ma_2^2\ln(a_3/a_2)}
\end{align}
for $\ln(a_3/a_2)\to+\infty$.

\begin{figure}[t]
\includegraphics[width=0.9\columnwidth,clip]{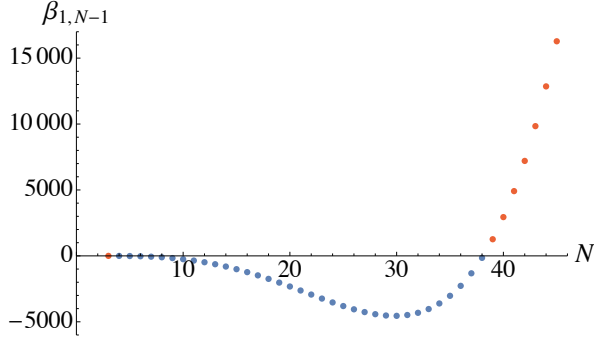}
\caption{\label{fig:beta}
Scattering length $\alpha_{1,N-1}$ between an atom and an $(N-1)$-body bound state induced by an infinitesimal three-body attraction in the form of $\beta_{1,N-1}\equiv a_2\ln(a_3/a_2)/(\sqrt3\pi\alpha_{1,N-1})$.
It turns out to be positive for $N=3$ and $N\geq39$ but negative for $4\leq N\leq38$ as indicated by the different colors.}
\end{figure}

Similarly, regarding the scattering state consisting of an atom with momentum $k$ and an $(N-1)$-body bound state at rest, its wave function in the domain of $x_1<x_2<\dots<x_N$ is provided by
\begin{align}\label{eq:scattering-state}
\Psi_{1,N-1}(\x) &= \sum_{j=1}^N\frac{(N-2-ika_2)(N-ika_2)}{(N-2j-ika_2)(N-2j+2-ika_2)} \notag\\
&\quad \times \frac{e^{ikx_j}}{\sqrt{NL}}\Psi_{N-1}(\x\backslash\{x_j\}),
\end{align}
where $\x\backslash\{x_j\}$ refers to $\x$ with $x_j$ excluded.
Because the wave function factorizes as $\Psi_{1,N-1}(\x)\to\frac{e^{ikx_j}}{\sqrt{NL}}\Psi_{N-1}(\x\backslash\{x_j\})$ at a large separation $x_j\ll\x\backslash\{x_j\}$, the scattering length between the atom and the $(N-1)$-body bound state is divergent, i.e., noninteracting~\cite{Castin:2001,Yurovsky:2006,Petrov:2012}.
Then, the expectation value of the renormalized three-body interaction energy $V_3=[\bar u_3(\kappa)/m]\sum_{1\leq i<j<k\leq N}\delta(x_{ij})\delta(x_{jk})$ with respect to the wave function in Eq.~(\ref{eq:scattering-state}) at $k\to0$ is found to be
\begin{align}
\lim_{k\to0}\<V_3\>_{1,N-1} = \Delta E_{N-1} - \frac{N}{(N-1)m\alpha_{1,N-1}L},
\end{align}
where the leading term is just the binding-energy shift in Eq.~(\ref{eq:ground_N-body}) but the subleading term reflects the interaction between the atom and the $(N-1)$-body bound state induced by an infinitesimal three-body attraction.
The extracted scattering length $\alpha_{1,N-1}\equiv a_2\ln(a_3/a_2)/(\sqrt3\pi\beta_{1,N-1})$ is plotted in Fig.~\ref{fig:beta} and turns out to be positive for $N=3$ and $N\geq39$ but negative for $4\leq N\leq38$, which correspond to the attractive and repulsive interactions between the atom and the $(N-1)$-body bound state, respectively.
Therefore, they in the former case with $\alpha_{1,N-1}\gg a_2$ constitute another $N$-body bound state induced by the infinitesimal three-body attraction, whose binding energy measured from the threshold at $E=E_{N-1}$ reads
\begin{align}\label{eq:excited_N-body}
-\frac{N}{2(N-1)m\alpha_{1,N-1}^2} = -\frac{3\pi^2N\beta_{1,N-1}^2}{2(N-1)ma_2^2\ln^2(a_3/a_2)}
\end{align}
for $\ln(a_3/a_2)\to+\infty$.
The values of $\beta_{1,N-1}$ for some selected $N$ are presented in Table~\ref{tab:beta}.

\begin{table}[t]
\caption{\label{tab:beta}
Values of $\beta_{1,N-1}$ for some selected boson numbers $N$.}
\begin{ruledtabular}
\begin{tabular}{rrlrr}
$N$ & \multicolumn{1}{c}{$\beta_{1,N-1}$} && $N$ & \multicolumn{1}{c}{$\beta_{1,N-1}$} \\[2pt]\hline
3 & $2/9\phantom{00}$ && 20 & $-2.32241\times10^3$ \\
4 & $-3\phantom{/000}$ && 30 & $-4.54773\times10^3$ \\
5 & $-184/15\phantom{0}$ && 40 & $2.94072\times10^3$ \\
6 & $-275/9\phantom{00}$ && 50 & $4.06680\times10^4$ \\
7 & $-19162/315$ && 100 & $2.32605\times10^6$ \\
8 & $-1589/15\phantom{0}$ && 200 & $6.36300\times10^7$ \\
9 & $-22744/135$ && 300 & $3.99017\times10^8$ \\
10 & $-6269/25\phantom{0}$ && 400 & $1.43180\times10^9$
\end{tabular}
\end{ruledtabular}
\end{table}

Beyond the limit of infinitesimal three-body attraction, the binding energies of $N$ bosons are to be determined by generalizing Eqs.~(\ref{eq:coupled_psi}) as
\begin{widetext}
\begin{subequations}
\begin{align}
& \left[\frac{a_2}{2} - \frac1{2\sqrt{\kappa^2+\frac14\left(\sum_{i=3}^Np_i\right)^2+\sum_{i=3}^N\frac{p_i^2}{2}}}\right]
\tilde\psi_2(\p\backslash\{p_1,p_2\}) \notag\\
&= \int\!\frac{dp_2}{2\pi}\frac1{\kappa^2+\frac12\left(\sum_{i=2}^Np_i\right)^2+\sum_{i=2}^N\frac{p_i^2}{2}}
\left[\sum_{1\leq i<j\leq N}^{(i,j)\neq(1,2)}\tilde\psi_2(\p\backslash\{p_i,p_j\})
+ \sum_{1\leq i<j<k\leq N}\tilde\psi_3(\p\backslash\{p_i,p_j,p_k\})\right]_{p_1\to-\sum_{i=2}^Np_i}
\end{align}
and
\begin{align}\label{eq:three-body_N}
& \left[\frac1{\sqrt3\pi}\ln\!\left(a_3\sqrt{\kappa^2+\frac16\left(\sum_{i=4}^Np_i\right)^2+\sum_{i=4}^N\frac{p_i^2}{2}}\right)\right]
\tilde\psi_3(\p\backslash\{p_1,p_2,p_3\}) \notag\\
&= \int\!\frac{dp_2dp_3}{(2\pi)^2}\frac1{\kappa^2+\frac12\left(\sum_{i=2}^Np_i\right)^2+\sum_{i=2}^N\frac{p_i^2}{2}}
\left[\sum_{1\leq i<j\leq N}\tilde\psi_2(\p\backslash\{p_i,p_j\})
+ \sum_{1\leq i<j<k\leq N}^{(i,j,k)\neq(1,2,3)}\tilde\psi_3(\p\backslash\{p_i,p_j,p_k\})\right]_{p_1\to-\sum_{i=2}^Np_i}.
\end{align}
\end{subequations}
\end{widetext}
While elaborate analyses of these coupled integral equations are deferred to a future work, we note that Eq.~(\ref{eq:three-body_N}) without $\tilde\psi_2$ was solved numerically for $N=4$ in the absence of the two-body interaction $a_3/a_2=0$~\cite{Sekino:2018}.
Here three four-body bound states (tetramers) were found with their binding energies provided by $\kappa=873.456/a_3$, $11.7181/a_3$, and $1.45739/a_3$.
On the other hand, in the opposite limit $a_3/a_2\to+\infty$ where the three-body attraction is infinitesimal, we find above that there exists only one tetramer state with its binding energy $\kappa\to\sqrt{10}/a_2$.
Therefore, the bound-state spectrum of four or more bosons as a function of $a_3/a_2$ is rather nontrivial and should be elucidated in the future work.

\section{Conclusion}
In this work, we studied bosons in one dimension with weak two-body and three-body interactions, predicted the existence of two trimer states when both interactions are attractive, and determined their binding energies as universal functions of the two-body and three-body scattering lengths.
We also showed that an infinitesimal three-body attraction induces an excited bound state only for 3, 39, or more bosons.
Because the effective three-body attraction inevitably arises by confining weakly interacting bosons into lower dimensions, our findings herein have direct relevance to a broad range of quasi-one-dimensional systems realized with ultracold atoms~\cite{Bloch:2008,Mora:2005,Petrov:2012,Pricoupenko:2018}.
In particular, when $a_\mathrm{3D}<0$ and $|a_\mathrm{3D}|\ll l_\perp$, the $N$-body to dimer binding-energy ratios predicted from Eqs.~(\ref{eq:coupling}), (\ref{eq:scattering_length}), (\ref{eq:ground_N-body}), and (\ref{eq:excited_N-body}) read
\begin{align}
\frac{E_N}{E_2} = \frac{E_N^\mathrm{(MG)}}{E_2}
+ \frac{4N(N^2-1)(N^2-4)\ln(4/3)}{15}\left(\frac{a_\mathrm{3D}}{l_\perp}\right)^2
\end{align}
for the ground state and
\begin{align}
\frac{E_N^*}{E_2} = \frac{E_{N-1}}{E_2}
+ \frac{72N\beta_{1,N-1}^2\ln^2(4/3)}{N-1}\left(\frac{a_\mathrm{3D}}{l_\perp}\right)^4
\end{align}
for the excited state with $N=3$ or $N\geq39$,%
\footnote{The existence and binding energies of these bound states for $N=3$ in quasi-one dimension were first presented in Ref.~\cite{Pricoupenko:talk}.}
which may be observable in ultracold atom experiments.

\acknowledgments
This work was supported by JSPS KAKENHI Grants No.~JP15K17727 and No.~JP15H05855.

\end{document}